# Self-amplified Cherenkov radiation from a relativistic electron in a waveguide partially filled with a laminated material

**L Sh Grigoryan**[1,3], **A R Mkrtchyan**[1], **H F Khachatryan**[1], **S R Arzumanyan**[1] and **W Wagner**[2]

[1]NAS Institute of Applied Problems in Physics, 25 Hr. Nersessian Str.,0014 Yerevan, Armenia
[2]Helmholtz-Zentrum Dresden-Rossendorf, POB 510119, 01314 Dresden, Germany

E-mail: levonshg@mail.ru

**Abstract.** The radiation from a relativistic electron uniformly moving along the axis of cylindrical waveguide filled with laminated material of finite length is investigated. Expressions for the spectral distribution of radiation passing throw the transverse section of waveguide at large distances from the laminated material are derived with no limitations on the amplitude and variation profile of the layered medium permittivity and permeability. Numerical results for layered material consisting of dielectric plates alternated with vacuum gaps are given. It is shown that at a special choice of problem parameters, Cherenkov radiation generated by the relativistic electron inside the plates is self-amplified. The visual explanation of this effect is given and a possible application is discussed.

**Keywords:** Waveguide, periodic medium, relativistic particle, Cherenkov radiation

## 1. Introduction

The presence of matter essentially influences the radiation from a relativistic particle: Cherenkov radiation, transition radiation etc. [1-9]. The effects of interest arise in stratified and periodic media (see e.g. [2,9]). The interfaces of media are widely used to control the radiation flow emitted by various systems. The conjuncture of three factors: the influence of material, the periodicity of medium and the presence of interfaces, provides a promising research area for development of high power controlled radiation sources.

An investigation of such a kind was carried out for a charged particle inside an infinite waveguide completely filled with a layered (spatially periodic) medium [10]. However, the case of Cherenkov radiation (CR) generation was not considered in [10] and has been partially investigated in [11,12] as the emission of CR from relativistic particles inside an infinitely long waveguide filled with semi-infinite layered medium. It was shown that CR may self-amplify due to the presence of a waveguide and of a periodical medium. In the present work a more realistic problem of filling an infinitely long waveguide with periodic medium of finite length is solved, and visual explanation of obtained numerical results is given.

---

[3] To whom any correspondence should be addressed.

In [13] the authors reported the first direct observation of narrow-band terahertz coherent CR driven by subpicosecond electron bunches travelling along the axis of a hollow cylindrical dielectric-lined waveguide (see also [14]). The results obtained both in [11,12] and in the present work permit an increase in the power of coherent CR from subpicosecond electron bunches observed in [13,14].

**2. Formulation of the problem**

Let us consider a relativistic particle of $q$ charge uniformly moving with v velocity along the axis of an ideal and infinitely long cylindrical waveguide of $R$ radius. We shall assume that the finite part of the waveguide is filled with a laminated material that weakly absorbs the radiation. The case of a waveguide filled with a stack of three plates is shown in Fig.1.

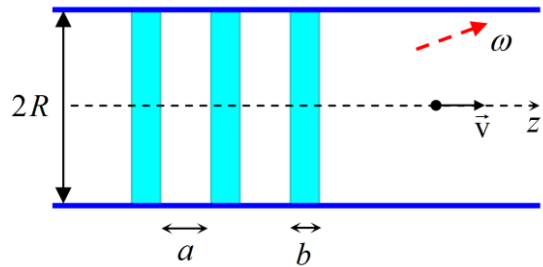

**Figure 1.** The radiation from a charged particle moving along the axis of cylindrical waveguide filled with a stack of three plates.

Now direct the Z axis of the cylindrical system of coordinates $r, \varphi, z$ along the waveguide axis and assume that the charge travels from vacuum ($z < z_1$) to the laminated material ($z_1 < z < z_2$) and then from laminated material again to vacuum ($z > z_2$).

We shall assume that the permittivity $\varepsilon_1$ and permeability $\mu_1$ of the layered medium are independent of the transverse coordinates $r, \varphi$, but alternate along the direction of particle motion according to an arbitrary law:

$$\varepsilon_1(z+l) = \varepsilon_1(z), \qquad \mu_1(z+l) = \mu_1(z) \qquad (1)$$

($l$ is the period of layered medium). The permittivity $\varepsilon$ and permeability $\mu$ of the system may be written in the form

$$\varepsilon(z) = \varepsilon_1(z), \qquad \mu(z) = \mu_1(z) \qquad \text{when} \quad z_1 < z < z_2 \qquad \text{(range 1)}$$
$$\varepsilon(z) = 1, \qquad \mu(z) = 1 \qquad \text{when} \quad z < z_1 \text{ or } z > z_2 \qquad \text{(range 2)}. \qquad (2)$$

We have studied the energy of radiation

$$W = \sum_{n=1}^{\infty} \int_{\omega_n}^{\infty} I_n(\omega) d\omega = \sum_n W_n \qquad (3)$$

emitted during the whole period of the particle motion and passing through the transverse section of waveguide in the $z = z_3$ point at large distances from the laminated material ($z_3 - z_2 \gg l$). In (3) $I_n$ and $W_n$ are the spectral distribution and the energy of radiation at the $n$-th mode of waveguide respectively.

The aim of the paper is the identification, analysis and clear explanation of cases when the waveguide and the periodical structure of layered material jointly strongly influence the spectral distribution of CR from a particle.

## 3. The stages of analytical calculations

Taking into account the azimuthal symmetry of the problem and making the Fourier transform

$$f_\omega = \frac{1}{2\pi}\int f(t)\exp(i\omega t)\mathrm{d}t, \tag{4}$$

one may reduce the set of Maxwell equations to the solution of one equation [10]

$$[\varepsilon\frac{\mathrm{d}}{\mathrm{d}z}\left(\frac{1}{\varepsilon}\frac{\mathrm{d}}{\mathrm{d}z}\right)+\frac{\omega^2}{c^2}\varepsilon\mu-\frac{\alpha_n^2}{R^2}]A_n = \frac{\varepsilon}{\mathrm{v}}\frac{\mathrm{d}}{\mathrm{d}z}\left(\frac{1}{\varepsilon}\exp(i\omega z/\mathrm{v})\right)-i\frac{\omega}{c^2}\varepsilon\mu\exp(i\omega z/\mathrm{v}), \tag{5}$$

that determines the longitudinal component $E_{z\omega}$ of electric field strength by means of formula

$$\varepsilon E_{z\omega}(r,z) = \sum_{n=1}^{\infty}\frac{2q}{\pi R^2 J_1^2(\alpha_n)}J_0(\tfrac{\alpha_n}{R}r)A_n(z). \tag{6}$$

Here $\alpha_n$ is the $n$-th root of the zeroth order Bessel function: $J_0(\alpha_n)=0$, and the coefficient $2q/\pi R^2 J_1^2(\alpha_n)$ is introduced for ease in treatment ($J_1(x)$ is the first order Bessel function). In these expressions $\varepsilon\equiv\varepsilon_\omega$ and $\mu\equiv\mu_\omega$ are Fourier transforms of the permittivity and permeability (with due regard for the frequency dispersion, the contribution of spatial dispersion being assumed negligibly small). We suppose that $\varepsilon_{1\omega}$ and $\mu_{1\omega}$ are complex valued functions for taking into account the absorption of radiation by the laminated material. Thus, (5) is the main equation of the problem under consideration.

In a hollow part of waveguide ($z > z_2$) the plane waves propagate:

$$A_n(z) = A_n^q(z) + \frac{i}{\omega}a_2\exp(ik_2 z), \qquad k_2 = \sqrt{\omega^2/c^2 - \alpha_n^2/R^2}. \tag{7}$$

Here the first summand describes the known field of charge inside the hollow waveguide, and the 2-nd summand ($a_2$ is a dimensionless quantity) describes the free field (the radiation field), if the following condition is observed

$$\omega > \omega_n = \alpha_n c/R. \tag{8}$$

Far from the interface $z = z_2$ of the laminated material with vacuum the field of produced radiation is «separated» from the field of charged particle and freely propagates inside the waveguide along the positive sense of the axis Z. The spectral distribution of this radiation over the cross section of waveguide on its $n$-th bandwidth is determined by means of well-known formula

$$I_n(\omega) = \frac{4q^2}{\pi\omega}\frac{k_2|a_2|^2}{\alpha_n^2 J_1^2(\alpha_n)}. \tag{9}$$

For calculation of amplitude $a_2$ it is sufficient to compare (7) with the complete solution of (5) that is valid for all $-\infty < z < \infty$. Such solution may be obtained using the method of Green functions. It is convenient as it is based on the solution of homogeneous equation corresponding to (5). Inside the laminated material that equation transforms to

$$[\varepsilon_1\frac{\mathrm{d}}{\mathrm{d}z}\left(\frac{1}{\varepsilon_1}\frac{\mathrm{d}}{\mathrm{d}z}\right)+\frac{\omega^2}{c^2}\varepsilon_1\mu_1-\frac{\alpha_n^2}{R^2}]G(z)=0. \tag{10}$$

According to the Bloch theorem an arbitrary solution of (10) may be presented as a superposition

$$G(z) = c_1 B_1(z) + c_2 B_2(z) \qquad \text{(range 1)} \tag{11}$$

of Bloch waves

$$B_m(z) = w_m(z)\exp(i\mu_m k_1 z), \quad \text{where } w_m(z+l) = w_m(z) \text{ and } \mu_m = \begin{cases} -1 & \text{when } m=1 \\ +1 & \text{when } m=2 \end{cases} \tag{12}$$

(travelling waves $\exp(\mp i k_1 z)$, modulated in the period of the layered medium). Here the quasiwave-number $k_1(\omega)$ is determined with an accuracy to the sign. To fix the sign we shall assume that

$$k_1 = k_1' + i k_1'', \qquad \text{where} \qquad \omega k_1''(\omega) \geq 0 \tag{13}$$

(the imaginary part $k_1'' \neq 0$, because the allowance for absorption of radiation by the material of layered medium is made).

It is known (see *e.g.* [15]) that one can represent the Bloch functions as sums of two arbitrary independent solutions $u_{1;2}(z)$ of equation (10):

$$B_m(z) = p_{1;m} u_1(z) - p_{2;m} u_2(z). \tag{14}$$

We shall need the following expressions

$$p_{1;m} = u_2(l+0) - \delta^{\mu_m} u_2(+0)$$
$$p_{2;m} = u_1(l+0) - \delta^{\mu_m} u_1(+0) \qquad \text{where} \qquad \delta = \exp(i k_1 l) \tag{15}$$

for expansion coefficients in (14). Using some arbitrariness in the choice of $u_{1;2}(z)$ we can simplify the calculations.

## 4. The general formula and its special cases

Omitting the intermediate calculations we now give the resultant expression

$$a_2 = \frac{\omega}{v k_2 \gamma_-} \left[ \eta_2 \xi_2 - \eta_1 \xi_1 + (\gamma - \sigma)(\varepsilon_1^{-1} - 1) + \tfrac{1}{2}(1 - \tfrac{v^2}{c^2}) \left( \frac{\gamma_- - 2\sigma}{1 - k_2 v/\omega} + \frac{1 - \delta^{2N}}{1 + k_2 v/\omega} \right) \right] \tag{16}$$

for the case when the first range contains $N = 1;2;3...$ periods of layered medium: $z_2 - z_1 = N \cdot l$. In (16)

$$\xi_m = i\omega/v \int_{+0}^{l+0} (\varepsilon_1^{-1} - \mu_1 v^2/c^2 + i v \dot{\varepsilon}_1 / \omega \varepsilon_1^2) u_m \exp(i\omega z/v) dz \tag{17}$$

and then

$$\eta_m = p_{m;2} B_{2+}^{-1} \frac{1 - \Delta_2^{-N}}{1 - \Delta_2} - p_{m;1} B_{1+}^{-1} \delta^{2N} \frac{1 - \Delta_1^{-N}}{1 - \Delta_1},$$

$$\gamma = B_1/B_{1+} - \delta^{2N} B_2/B_{2+}, \qquad \gamma_- = B_{1-}/B_{1+} - \delta^{2N} B_{2-}/B_{2+},$$

$$\sigma \cdot \Delta_1^N = B_1/B_{1+} - B_2/B_{2+}, \qquad \Delta_m = \delta^{\mu_m} \exp[i\omega l/v] \qquad B_{m\pm} = B_m \pm \dot{B}_m / i k_2 \varepsilon_1. \tag{18}$$

Functions $\varepsilon_1(z)$, $B_m(z)$ and $B_{m\pm}(z)$ in (16) and (18) shall be taken in $z = 0$ point, the point over the function implies the differentiation with respect to its argument. In (16) there are no limitations on the amplitude and variation profile of $\varepsilon_1(z)$ and $\mu_1(z)$.

Equations (16) and (18) are simplified when the waveguide is filled with weakly absorbing semi-infinite layered medium ($z_1 \to -\infty$), as $\delta^N \to 0$ when $N \to \infty$. An appropriate expression for $a_2$ (in other equivalent form) is given in [11]. In the particular case of $\varepsilon_1(z)$, $\mu_1(z) = const$ the equation (16) describes the radiation field of particle passing through the dielectric plate inside the waveguide [16].

## 5. Numerical results based on (16)

Below we shall consider a special, but highly advantageous case of a laminated material consists of $N$ plates (each of $b$ length) interleaved with vacuum gaps (the length of a gap is $a$, Fig.1). The corresponding expression for $a_2$ is given in [17], but here the case of CR has not been investigated. Now investigate CR generated inside a waveguide filled with a stack of plate

**Table 1.** The parameters of the problem.

| Curve | $a/R$ | $b/R$ |
|---|---|---|
| A | 308.6 | 13.79 |
| B | - | 3x13.79 |
| C | 15.43 | 13.79 |

In Fig. 2 three curves of the spectral distribution $I_n(\omega)$ of CR from an electron on the 3$^{rd}$ mode of waveguide are shown. The thicknesses of plates and gaps in between are given in Table 1 and the permittivity and permeability of the material of plates are

$$\varepsilon_b = \varepsilon_b' + i\varepsilon_b'' = 1.3 + 0.005i, \qquad \mu_b = 1 \qquad (19)$$

respectively (we assume that $\varepsilon_b$ is a weak function of the frequency in the wavelength range under consideration). The energy of electron is 1.2 MeV so that the Cherenkov condition is satisfied: $v > c/\varepsilon_b'^{1/2}$.

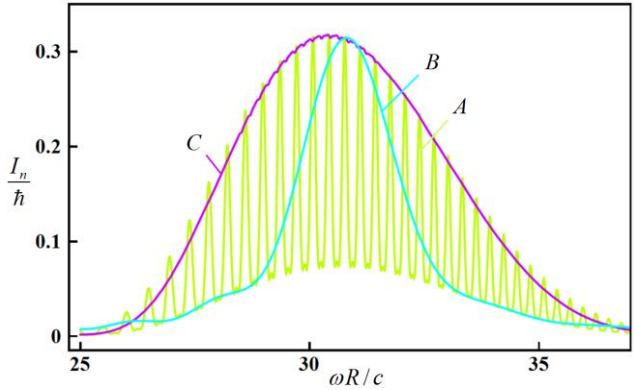

**Figure 2.** The spectral distribution of CR from an electron at the 3$^{rd}$ mode of waveguide.

In the case of curve A the waveguide is filled with a stack of $N = 3$ plates. The thicknesses of vacuum gaps and plates are selected as is given in Table 1 (the row A). It is clear that the pulses of CR generated by the particle in different plates shall superimpose and, hence, the radiation spectrum has to be an oscillating curve. Now compare this curve with the curve B, when all the three plates are so close to each other that they merge and form one thick plate ($N = 1$, one plate filling of the waveguide, the row B in Table 1). As was to be expected, the oscillation of spectrum disappears. And what is clear, the total energy of radiation for the curves A and B is the same as, in general, it does not change at interference:

$$W_3^A \approx W_3^B \approx 0.95 c\hbar/R. \qquad (20)$$

Add besides, that the location of the maximum in CR spectrum is described by the known simple formula:

$$\omega_n^{max} \approx \frac{\alpha_n v}{R\sqrt{\varepsilon_b' v^2/c^2 - 1}}. \qquad (21)$$

It is quite clear. It is unique that another situation is possible.

Now return to the case when the waveguide filling is a stack of three plates and select another special value of the thickness of vacuum gaps between the plates (Table 1, the row C). The corresponding spectral distribution of CR is also depicted in Fig.2 (the curve C). It is seen that at the transition $A \to C$ the oscillations practically disappear. Moreover, the total energy of radiation (the surface under the curve C) is almost twice larger than that for the curve A or B:

$$W_3^C = 1.76 c\hbar/R. \qquad (22)$$

In Fig.3 it is shown the radiation spectra at the first five modes of waveguide (the number of mode is seen beside the curves B and C (see Table 1). The curves B are given for comparison (waveguide filled with one thick plate).

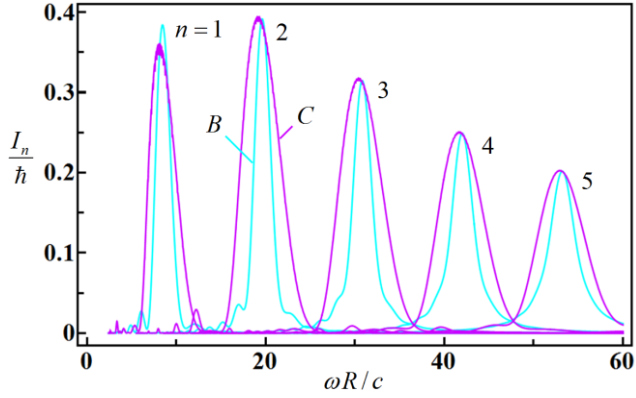

**Figure 3.** The radiation spectra at the first five modes of waveguide.

From Fig. 3 it follows that at the transition $B \to C$ the total energy of radiation increases at all five waveguide modes.

## 6. Visual explanation
In Fig.4 two neighboring plates inside the waveguide are shown.

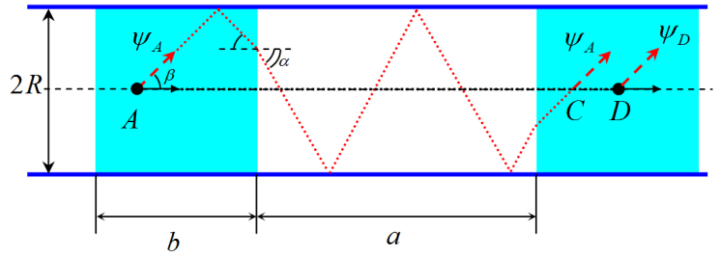

**Figure 4.** In case of $C \equiv D$ the pulse $\Psi_D$ of CR generated by the particle in the vicinity of point D is «superimposed» on the pulse $\Psi_A$ of CR generated by the particle in the vicinity of point A prior to that.

Now let us consider the instant when the relativistic particle was in the left plate and observe $\Psi_A$ pulse of CR, generated by a particle in the immediate vicinity of point A, that is emitted at a specific angle β (angle of CR) with respect to the waveguide axis (dashed line). At propagation this pulse is found in the right plate and at some instant of time crosses the particle trajectory (the waveguide axis) in a certain point C at the same angle β. By this time the relativistic particle emits $\Psi_D$ pulse of CR in the immediate vicinity of some (generally) other point D (see Fig.4). However, if the following two equations

$$\frac{a+b}{v} = \frac{a}{c\cos\alpha} + \frac{b\sqrt{\varepsilon'_b \mu'_b}}{\cos\beta}, \qquad \frac{a\operatorname{tg}\alpha + b\operatorname{tg}\beta}{2R} = 4s \qquad (23)$$

are satisfied (α is an angle giving the direction of CR with respect to the waveguide axis in the vacuum between the plates, $s$ is the integer), then the exceptional situation arises:

$$CD = 0 \quad \text{and} \quad AC = a+b, \qquad (24)$$

when, first, the point D coincides with the point C and, second, this «joined» point C-D in the right-side plate is located where in the left-side plate the point A is located. For this reason $\Psi_D$ pulse emitted at the same angle β in the vicinity of point D will "superimpose" on $\Psi_A$ pulse emitted earlier in the 1st plate. As a result of interference the pulses may be either suppressed or amplified depending on the specific value of phase difference. The factor 4 (see e.g. [2]) in the right-hand side of the second equation of (23) insures the fulfillment of the condition of pulses constructive superposition.

Summarizing one concludes that if equations (23) are met, then the Cherenkov waves are generated inside the plate and at the same time constructively interfere with those emitted by the particle in the preceding plates. In other words, the in-phase superposition of CR pulses takes place in the zone of radiation formation. The in-phase superposition of electromagnetic oscillations at point C-D entails the total field increase in the zone of radiation formation. The force that brakes the motion of particle will also increase and the extra work of external force compelling the particle to move uniformly along the waveguide axis will be spent for generation of more powerful CR.

The thicknesses of vacuum gaps and plates for the curves C in Figs.2,3 (see Table 1, the row C) have been determined from (23) at $s=1$. This fact explains the reason of CR amplification (more precisely, self-amplification of CR) at all five waveguide modes at the transition from the waveguide filled with one thick plate (curves B in Figs.2,3) to the waveguide filled with a stack of shorter plates, obtained from this long plates by its equal sharing (curves C in Figs.2,3).

## 7. Discussion

In 2009 J.B. Rosenzweig with co-authors reported [13] the fist direct observation of narrow-band terahertz coherent CR driven by a subpicosecond electron bunch traveling along the axis of a hollow cylindrical dielectric-lined waveguide. The measurements indicate a peak power of 150 kW. In 2011 J.B. Rosenzweig with co-authors implement structure variations that gave improved result [14].

We propose to improve the experimental setup of [13,14]. The hollow cylindrical dielectric inside the waveguide shall be cut in *N* identical parts (tubes) and then a periodic structure consisting of these tubes alternated with vacuum gaps shall be combined. If the length of each vacuum gap between the tubes is selected correctly, then (according to the preliminary calculations [12]) we expect nearly a *N*-fold increase in the peak power of coherent CR from electron bunches.

## 8. Conclusions

In this paper the radiation from a relativistic particle moving along the cylindrical waveguide axis has been investigated under the assumption that the waveguide is filled with a laminated material of finite length, the permittivity $\varepsilon_1(z)$ and permeability $\mu_1(z)$ of which are periodic functions of arbitrary shape.

1. Expression (16) for calculation of the spectral distribution (9) of radiation passing throw the transverse section of waveguide in vacuum (at large distances from the laminated material) have been derived with no limitations on the amplitude and variation profile of $\varepsilon_1(z)$ and $\mu_1(z)$.
2. Numerical results for CR from a relativistic electron uniformly moving along the axis of cylindrical waveguide filled with a stack of dielectric plates alternated with vacuum gaps have been investigated (Section 5).
3. The values of the problem parameters, for which the waveguide and the periodical structure of laminated material jointly strongly influence the spectral distribution of CR from an electron, have been investigated.
4. It has been shown that due to the existence of vacuum gaps between the plates the power of CR from an electron may be many times greater than that produced, when the plates in the stack are so close that one thick plate is formed (waveguide filled with a solid dielectric without vacuum gaps). The visual explanation of this effect of self-amplification of CR have been given (Section 6).
5. It was proposed to use this effect for amplification of the power of coherent CR from subpicosecond electron bunches observed by J.B. Rosenzweig with co-authors (Section 7).

**Acknowledgments**
One of the authors (L.Sh.G.) is thankful to A. P. Potylitsyn for stimulating discussion and encouragement, N. F. Shulga, B. M. Bolotovskii for valuable comments and R. Chehab for the interest to these investigations.